# Measures of Ten Sco Doubles and the Determination of Two Orbits.

Matthew James[1], Meg Emery[2], Graeme L. White[3], Roderick Letchford[4], Stephen Bosi[5].

1. University of New England, Armidale, NSW, Australia; m.b.james27@gmail.com
2. Kildare Catholic College, Wagga Wagga, NSW, Australia; megcolemery@gmail.com
3. Centre for Astronomy, University of Southern Queensland, Toowoomba, QLD, Australia; graemewhiteau@gmail.com
4. Vianney College Seminary, Wagga Wagga, NSW, Australia; rodvianney@yahoo.com.au
5. University of New England, Armidale, NSW, Australia; sbosi@une.edu.au

**Abstract**:

We present measures for 10 pairs in the constellation of Scorpius using a C14 telescope, *Lucky Imaging*, and the *Reduc* software. The separations of Alpha Centauri AB, as determined from the orbital elements of Pourbaix and Boffin (2016), were used as an image scale and position angle calibrator.

Our internal uncertainties are ~0.06 arcsec in $\rho$ and ~0.06 degree in PA. There is excellent agreement with historic data extrapolated to epoch of observation (~2018.53), and micro-arcsecond positions from the GAIA database where the differences are ~0.05 arcsec in $\rho$ and ~0.15 degrees in PA.

In addition, we present rectilinear elements for the 10 Sco pairs and orbital elements for two of them. Ephemera are given for these pairs based on both the rectilinear elements and the orbital elements.

## 1.    Introduction.

We present here the first of two papers that explore the limits of uncertainty that can be obtained using different techniques to determine standard separation, $\rho$, and position angles, PA. In this first paper we undertake lucky imaging measures of 10 pairs in the constellation of Scorpio (Sco) using drifting images, with the image scale and the camera's position angle calibrated against an accurate ephemeris of Alpha Centauri. A second paper (James *et al.,* in preparation for journal submission) will undertake a more detailed analysis of the accuracy of different applications of lucky imaging.

We present measures for these pairs and look for uncertainty through comparison with extrapolations of historic data and micro-arcsecond positions from the GAIA DR2 database. One method to determine if a double star system is a visual double or a binary system is to observe the relative motions between the primary and secondary component over a period of time. The trend can be used to differentiate between orbital or rectilinear motion. Section 5 looks at the motion of the pairs over time utilising the historic record, and Section 6 determines the rectilinear motion of the pairs following the method of Letchford, White and Ernest (2018a). For binary systems with very short orbital arcs the method developed by Letchford, White and Ernest (2018b) is used in Section 7 to determine grade 5 orbits for two of the pairs in this study.

**2.     Selection of Pairs.**

The objects in this study were chosen from the Carro Double Star Catalogue (Carro, 2013) that have a separation, ρ, larger than 4 arcseconds; a limit imposed by local seeing conditions.

The constellation Scorpius was specifically chosen since the constellation was near zenith at the time of observation. High elevations reduce the effects of air-mass and give better video captures.

Table 1 lists the stars that make up the 10 pairs. Here the WDS designation is given along with the WDS Discovery Code (*Disc*). Both are adapted from the Washington Double Star Catalog (WDS, Mason, *et al.,* 2001). The names/identifiers of the stars are from the SIMBAD database (Wegner, *et al.,* 2000), the ASCC database (Kharchenko, 2001) and from the DR2 release (Brown, *et al.*, 2018) of the GAIA astrometric mission (Prusti, *et al.*, 2016).

*Table 1*. *Modern Identifications of the Stars that make up the 10 Sco Pairs.*

|    | WDS        | Disc         | SIMBAD        | ASCC    | GAIA                |
|----|------------|--------------|---------------|---------|---------------------|
| 1  | 16029-2501 | BU 38 A,B    | TYC 6784-1420-1 | 1683158 | 6235913255305953536 |
|    |            |              | TYC 6784-1425-1 | 1683157 | 6235913255305954432 |
| 2  | 16095-3239 | BSO 11 A,B   | HD 144927     | 1778073 | 6035755719057732224 |
|    |            |              | TYC 7334-2610-1 | 1778075 | 6035755719057730816 |
| 3  | 16143-1025 | STF2019 AB,C | HD 145996     | 1401854 | 4344884406644977280 |
|    |            |              | BD-10  4276C  | 1401855 | 4344884406644973952 |
| 4  | 16195-3054 | BSO 12 A,B   | HD 146836     | 1778791 | 6037514800199297152 |
|    |            |              | HD 146835     | 1778788 | 6037514800213601024 |
| 5  | 16201-2003 | SHJ225 A,B   | V* V933 Sco   | 1587367 | 6244725050721030528 |
|    |            |              | HD 147009     | 1587364 | 6244725905417556992 |
| 6  | 16247-2942 | H N 39 A,B   | HD 147723     | 1684185 | 6038073970589665280 |
|    |            |              | HD 147722     | 1684184 | 6038073970589665536 |
| 7  | 16482-3653 | DUN 209 A,B  | HD 151315     | 1875678 | 5971596329361239808 |
|    |            |              | HD 151316     | 1875680 | 5971596260664472704 |
| 8  | 16510-3731 | HJ 4889 A,B  | HD 151771A    | 1875852 | 5971527094516942720 |
|    |            |              | HD 151771B    | 1875854 | 5971527094516943488 |
| 9  | 17290-4358 | DUN 217 A.B  | HD 158042     | 1978893 | 5958561447264080768 |
|    |            |              | CD-43 11741B  | 1978895 | 5958561447264078208 |
| 10 | 17512-3033 | PZ 5 A,B     | HD 162220     | 1788279 | 4056340704108946176 |
|    |            |              | CD-30 14802B  | 1788278 | 4056340704108937600 |

## 3. Observations.

*Equipment and Software.*

The telescope used to make the observations is the Bill Webster 14-inch Celestron Schmidt-Cassegrain Telescope located at the Kirby Observatory of the University of New England, Armidale, NSW, Australia. The telescope is equipped with a flip-mirror box which allows the user to switch between the camera and eyepiece. The camera, a ZWO ASI120MM-S USB 3.0 Monochrome CMOS, was used for its image resolution, temporal resolution, and the USB 3.0 download bandwidth. A red (approximating R) filter was used to reduce the effects of atmospheric distortion on the video captures.

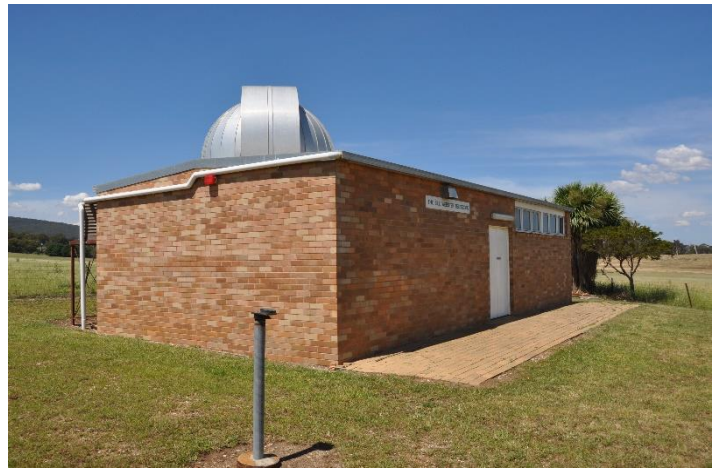

*The Kirby Observatory of the University of New England, Armidale, NSW, Australia. The dome covers the Bill Webster 14-inch Celestron Schmidt-Cassegrain Telescope used in this study.*

The video capture software used was *SharpCaps* version 3.1, and the analysis software used was *Reduc* version 5.36, provided by Florent Losse. *Reduc* allows for the rapid disposal of data (capture frames) resulting in a text file output of the X and Y coordinates of the primary and secondary star on the chip, which were used in Microsoft Excel 2016 to calculate the PA and ρ along with their formal uncertainties.

*Lucky Imaging*

The lucky imaging technique is utilized which is akin to high speed photography; high framerate and low exposure times. When used on astronomical objects, like double stars, the short exposure time (<100ms) has the effect of freezing the perturbed atmosphere, reducing image distortion and increasing the chance of obtaining higher quality images (Fried, 1977).

*Video Drift Method.*

Observations were made using the video drift method outlined by Nugent (2011). In brief, the pair is placed to the east side just outside of the video frame, the telescope's drive motor is deactivated and the video capture is then started, which results in the image of the pair drifting across the field of view of the camera (and out of the west side of the camera frame).

*Calibration – Image Scale and Position Angle.*

Calibration of the image scale and position angle was undertaken using the prominent southern hemisphere pair Alpha Centauri (α Cen). α Cen has been extensively studied (see White, Letchford and Ernest, 2018) over ~260 years and ~3.5 orbits. Precise orbital elements by Pourbaix and Boffin (2016) are available in the WDS Sixth Orbit Catalog, and these give precise predicted ρ and PA for the pair at the epoch of observation. Concurrent observations of α Cen underpin the calibration of the Sco measures presented here, and uncertainties in the calibration observations of α Cen contribute to the uncertainties in the measures presented below.

*Analysis.*

For each Sco pair, 5 AVI format video captures were taken using the video drift method. Each video capture was then reduced using all frames with ρ or PA outside of 2 standard deviation from the mean being removed. Further reduction was undertaken using Excel. Further details of the data analysis are given by James (2019).

*Image Scale for the Calibration of Separation.*

To determine the ρ for the 10 Sco pairs, a pixel per arcsecond ratio was calculated for the C14/ZWO ASI120MM-S telescope/camera based on the observations of α Cen; the image scale, $R_{px/as}$, was determined to be 5.664 pixels per arcsecond. Again further details of the data analysis are given in James (2019).

*Position Angle.*

The position angle was computed from the drift angle of the individual stars across the chip of the camera, the individual positions on each frame having been determined using *Reduc*. Again further details are given by James (2019).

### 4. Measures.

The measures for the 10 Sco objects observed are given in Table 2. The formal uncertainties in these measures are the uncertainty of the observations of the Sco pairs combined with the uncertainty in the calibration observations of α Cen (for the ρ). These uncertainties are the Standard Error in the Mean (SEM) of 5 independent observations.

Equation 1 was used to calculate the Standard Error in the Mean (SEM) uncertainty of the measures. This equation combines both the uncertainty in the observations and in the calibrator into a single SEM uncertainty. Here N is the number of observations of a particular pair; always N=5 for this paper. $\overline{M}$ is the average measure (either PA or ρ) of the N number of observations. Cal refers to the calibrator i.e. α Cen. $\sigma$ is the standard deviation (SD) of the measures of N observations.

*Equation 1. The equation used to calculate the standard error in the mean for PA and ρ*

$$SEM = \overline{M} \sqrt{\left(\frac{\sigma_M}{\overline{M}\sqrt{N_M}}\right)^2 + \left(\frac{\sigma_{cal}}{M_{cal}\sqrt{N_{cal}}}\right)^2}$$

*Table 2. Measures at Epoch for 10 Sco Pairs.*

|   | WDS | Disc | Epoch | Separation (arcsec) | SEM (arcsec) | PA (deg) | SEM (deg) |
|---|---|---|---|---|---|---|---|
| 1 | 16029-2501 | BU 38 | 2018.542 | 4.47 | 0.02 | 343.09 | 0.28 |
| 2 | 16095-3239 | BSO 11AB | 2018.526 | 7.64 | 0.03 | 83.61 | 0.02 |
| 3 | 16143-1025 | STF2019AB,C | 2018.542 | 22.32 | 0.10 | 153.02 | 0.09 |
| 4 | 16195-3054 | BSO 12AB | 2018.523 | 23.58 | 0.10 | 317.98 | 0.02 |
| 5 | 16201-2003 | SHJ225 | 2018.542 | 46.69 | 0.19 | 332.60 | 0.02 |
| 6 | 16247-2942 | H N 39 | 2018.526 | 4.00 | 0.02 | 359.24 | 0.05 |
| 7 | 16482-3653 | DUN 209AB | 2018.545 | 23.91 | 0.10 | 137.97 | 0.03 |
| 8 | 16510-3731 | HJ 4889 | 2018.526 | 6.76 | 0.03 | 4.22 | 0.06 |
| 9 | 17290-4358 | DUN 217 | 2018.526 | 13.46 | 0.06 | 167.84 | 0.02 |
| 10 | 17512-3033 | PZ 5AB | 2018.526 | 10.10 | 0.04 | 189.32 | 0.05 |

### 5. Historic Observations.

Historic positional measures have been obtained from the supplementary catalogues of the WDS. For the 10 pairs studied here, there are a total of 420 observations starting as early as 1783.23 (for the pair WDS 16201-2003).

Appendix 1 presents the ρ and PA for the 10 Sco pairs. Datapoints as orange squares were deemed to be outliers and rejected based on a subjective assessment of the trend. The green triangle datapoints are the measures from this work (taken from Table 2).

*Precession of Position Angles.*

All plots in Appendix 1 are for PA (and ρ) at the epoch and equinox of observation. The correction of the PA to bring them to a standard Equinox (say J2000) have not been applied. This precessional rotation of the frame is defined in Aitken (1935, p. 73) and Argyle (2004). As all pairs in the study are in close proximity (~$16^h$ $30^m$, -20º) the PA precession of each pair is approximately 0.55 degree per century in the sense that the PA is decreasing with time. A much smaller component of PA precession based on the proper motion of the primary star (Argyle, 2004) was ignored in this work, except for those in Appendices 2 & 3, Tables 6, 7, & 9.

*Uncertainties in Historic Measures.*

White, Letchford and Ernest (2018) have shown that the precision of historic observations of double stars has improved with epoch; from ~0.6 arcseconds to ~0.14 arcseconds in ρ, and ~0.74 degree to ~0.5 degree in PA, over our period of interest (~1800 to the present). This trend towards better quality data is also visible here as it is seen that the spread of data points around the trend line converges with increasing time.

*Fitted Trend Lines.*

Each plot in Appendix 1 has been fitted by an unweighted linear trend line and the fitted parameters are given in Table 3 along with the derived correlation coefficient, $R^2$.

For the fit to the separation, ρ, with Epoch plot, the fit is defined as

$$\text{Separation}, \rho = \mathbf{A} \times \mathbf{Epoch} + \mathbf{B}$$

and the fitted trend line for the Position Angle, PA, is

$$\text{Position Angle}, \text{PA} = \mathbf{C} \times \mathbf{Epoch} + \mathbf{D}$$

The fitted parameters **A**, **B**, **C** and **D** are presented in Table 3 along with the fitted correlation co-efficient, $\mathbf{R^2}$.

*Table 3. Linear fits Co-efficient to the Historic Data for 10 Sco Pairs.*

|   | WDS | Disc | Rho | | | PA | | |
|---|---|---|---|---|---|---|---|---|
|   |     |      | A | B | $R^2$ | C | D | $R^2$ |
| 1 | 16029-2501 | BU 38 | 0.00126 | 1.965 | 0.061 | -0.0605 | 465.907 | 0.896 |
| 2 | 16095-3239 | BSO 11AB | -0.00448 | 16.606 | 0.118 | -0.0148 | 113.872 | 0.300 |
| 3 | 16143-1025 | STF2019AB,C | 0.00908 | 4.589 | 0.316 | 0.0032 | 147.704 | 0.014 |
| 4 | 16195-3054 | BSO 12AB | -0.00432 | 31.821 | 0.057 | -0.0147 | 347.660 | 0.213 |
| 5 | 16201-2003 | SHJ225 | -0.00124 | 49.204 | 0.033 | -0.0027 | 338.412 | 0.067 |
| 6 | 16247-2942 | H N 39 | -0.01886 | 42.323 | 0.859 | 0.0503 | 256.409 | 0.617 |
| 7 | 16482-3653 | DUN 209AB | 0.00177 | 20.041 | 0.033 | -0.0563 | 251.380 | 0.937 |
| 8 | 16510-3731 | HJ 4889 | -0.00144 | 9.612 | 0.056 | -0.0090 | 22.915 | 0.201 |
| 9 | 17290-4358 | DUN 217 | -0.00449 | 22.343 | 0.375 | -0.0065 | 181.831 | 0.043 |
| 10 | 17512-3033 | PZ 5AB | 0.00023 | 9.671 | 0.002 | -0.0017 | 192.841 | 0.027 |

6. **Accuracy of the 10 Sco Measures.**

The measures for the 10 Sco pairs in Table 2 were now compared with two external measures (i) the historic measures extrapolated to the epoch of observation, and (ii) the position given in the GAIA DR2 catalogue which was precessed to the epoch of observation.

Using a fitted linear trendline extrapolated from the historic measures from Table 3, $\rho$ and PA are calculated for the epoch of observations and shown in Table 4 in column *History*.

The $\rho$ and PA for the pairs obtained at epoch from the GAIA DR2 data base are given in Table 4 under heading *GAIA*.

*Table 4. Comparison of Measures with (i) Extrapolated Historic Data and*
*(ii) GAIA positions and Proper Motions.*

|  | WDS | Disc | Rho | | | PA | | |
|---|---|---|---|---|---|---|---|---|
|  |  |  | This Work | History | GAIA | This Work | History | GAIA |
|  |  |  | (arcsec) | (arcsec) | (arcsec) | (deg) | (deg) | (deg) |
| 1 | 16029-2501 | BU 38 | 4.472 | 4.504 | 4.487 | 343.091 | 343.759 | 343.445 |
| 2 | 16095-3239 | BSO 11AB | 7.636 | 7.571 | 7.649 | 83.610 | 83.947 | 83.644 |
| 3 | 16143-1025 | STF2019AB,C | 22.320 | 22.924 | 21.625 | 153.021 | 154.089 | 155.359 |
| 4 | 16195-3054 | BSO 12AB | 23.579 | 23.109 | 23.498 | 317.976 | 318.046 | 318.180 |
| 5 | 16201-2003 | SHJ225 | 46.686 | 46.703 | 46.637 | 332.601 | 332.982 | 332.707 |
| 6 | 16247-2942 | H N 39 | 3.999 | 4.257 | 4.006 | 359.241 | 357.977 | 359.322 |
| 7 | 16482-3653 | DUN 209AB | 23.914 | 23.616 | 23.888 | 137.972 | 137.740 | 138.120 |
| 8 | 16510-3731 | HJ 4889 | 6.761 | 6.711 | 6.767 | 4.224 | 4.805 | 4.503 |
| 9 | 17290-4358 | DUN 217 | 13.457 | 13.278 | 13.433 | 167.837 | 168.680 | 168.040 |
| 10 | 17512-3033 | PZ 5AB | 10.101 | 10.136 | 10.110 | 189.323 | 189.498 | 189.580 |

The differences between the measures reported here (shown here as This Paper, *TP*) and those extrapolated from the historic data (shown as *Hist*) and the GAIA database are given in Table 5. Here units for differences in ρ are arcsecond, and degrees in PA respectively. There is excellent agreement between the measures of historic and GAIA values. The mean offset between the data sets (shown as *Average =*), and its formal uncertainty (shown as *SEM =*) are all self-consistent and consistent with the formal uncertainties quoted for the measures in Table 2. The average SEM in the offset of ρ is 0.04 arcseconds, and the average SEM in the offset of PA is 0.12 degrees.

This comparison does not extend to WDS 16143-1025. The brightest (primary) component observed in this work is WDS 16143-1025 AB, a very close pair separated by only 0.2 arcseconds. The positions, and proper motions, reported for components AC by both the HIPPARCOS and GAIA mission are inconsistent and no comparison has been made with the measures reported in Table 2.

*Table 5. Differences of Measures with (i) Extrapolated Historic Data and  
(ii) GAIA positions and Proper Motions.*

| WDS | Disc | Diff Rho | Diff PA | Diff Rho | Diff PA | Diff Rho | Diff PA |
|---|---|---|---|---|---|---|---|
| | | GAIA - TP | GAIA - TP | Hist - TP | Hist - TP | Hist - GAIA | Hist - GAIA |
| 16029-2501 | BU 38 | 0.015 | 0.353 | 0.032 | 0.668 | 0.017 | 0.314 |
| 16095-3239 | BSO 11AB | 0.012 | 0.034 | -0.065 | 0.337 | -0.077 | 0.303 |
| 16143-1025 | STF2019AB,C | | | | | | |
| 16195-3054 | BSO 12AB | -0.081 | 0.204 | -0.470 | 0.071 | -0.389 | -0.134 |
| 16201-2003 | SHJ225 | -0.050 | 0.106 | 0.017 | 0.380 | 0.066 | 0.274 |
| 16247-2942 | H N 39 | 0.007 | 0.080 | 0.258 | -1.264 | 0.252 | -1.344 |
| 16482-3653 | DUN 209AB | -0.027 | 0.148 | -0.299 | -0.232 | -0.272 | -0.379 |
| 16510-3731 | HJ 4889 | 0.006 | 0.279 | -0.050 | 0.582 | -0.056 | 0.303 |
| 17290-4358 | DUN 217 | -0.024 | 0.202 | -0.179 | 0.843 | -0.155 | 0.640 |
| 17512-3033 | PZ 5AB | 0.009 | 0.257 | 0.035 | 0.175 | 0.026 | -0.082 |
| | Average = | -0.015 | 0.185 | -0.080 | 0.173 | -0.065 | -0.012 |
| | SEM = | 0.011 | 0.034 | 0.071 | 0.210 | 0.063 | 0.195 |

## 7. Rectilinear Motion.

The motion of the components of a double star may be characterised as a rectilinear motion of the secondary relative to the primary star. Rectilinear motion is usually visualized as a straight line on a Cartesian plot where the primary star is the origin (0,0) position.

Such descriptions are an important tool in distinguishing between optical doubles and physical binaries since it is the variations from linearity that allows a sensitive identification of a Keplerian system.

As stated above, White, Letchford and Ernest (2018) have shown the precision of historic observations of double stars to be ~0.14 arcsec in $\rho$, and ~0.5 degree in PA, at best, for recent measures, and where the uncertainties for early measures are larger (~0.6 arcsec and ~74 degree). These uncertainties are dwarfed by the precisions of the HIPPARCOS and GAIA spacecraft (milli-arcsecond and micro-arcsecond respectively) and their inclusion in the rectilinear analysis presented here would not contribute to the accuracy of that analysis. The rectilinear analysis there is based only on the HIPPARCOS and GAIA positions and the historic measures are shown in Appendix 2 only for completeness, as are the measures from Table 2.

The rectilinear plots for the 10 Sco pairs are presented in Appendix 2.

Table 6 gives the Rectilinear Elements for the 10 pairs, where the column headings, x0, xa, y0, ya, t0, θ0 and ρ0 are defined in Letchford, White and Ernest, 2018a.

*Table 6. Rectilinear Elements for 10 Sco pairs.*

|   | WDS | x0 (DE0) +/- | xa +/- | y0 (RA0) +/- | ya +/- | t0 +/- | theta0 +/- | rho0 +/- | xb +/- | yb +/- | move +/- | x-inter | y-inter |
|---|---|---|---|---|---|---|---|---|---|---|---|---|---|
| 1 | 16029-2501 | 3.58 | -32 | -2.35 | -0.0050 | 2200 | 326.7 | 4.28 | 10.9 | 8.8 | 5.9 | 5.12 | -7.81 |
|   |            | 0.02 | 0.0001 | 0.07 | 0.0003 | 200 | 0.8 | 0.04 | 0.2 | 0.6 | 0.2 |   |   |
| 2 | 16095-3239 | 4.23 | 0.001079 | 3.6 | -0.00127 | 5000 | 40.0 | 5.6 | -1.34 | 10.2 | 1.67 | 7.28 | 8.59 |
|   |            | 0.07 | 2.23E-05 | 0.5 | 0.00014 | 900 | 4.0 | 0.3 | 0.04 | 0.3 | 0.11 |   |   |
| 3 | 16143-1025 | -14 | -0.005 | 15.4 | -0.004 | 700 | 133.0 | 21 | -11 | 19 | 6.2 | -30.80 | 28.60 |
|   |            | 2 | 0.002 | 2.5 | 0.002 | 1400 | 7.0 | 2 | 4 | 4 | 1.3 |   |   |
| 4 | 16195-3054 | 9 | 0.0026 | 4 | -0.0062 | -1000 | 23.0 | 10 | 12 | -3 | 6.7 | 10.99 | 26.16 |
|   |            | 4 | 0.0012 | 4 | 0.0012 | 700 | 20.0 | 4 | 2 | 3 | 1.1 |   |   |
| 5 | 16201-2003 | 3 | -0.00115 | -25.7 | -0.00013 | 35000 | 276.0 | 25.8 | 43.80 | -21 | 1.156 | 230.98 | -25.98 |
|   |            | 1 | 3.07E-05 | 1.5 | 4.33E-05 | 2000 | 2.0 | 1.5 | 0.06 | 0.09 | 0.004 |   |   |
| 6 | 16247-2942 | 0.598 | -0.02345 | 1.30 | 0.0090 | 2170 | 69.1 | 1.40 | 51.35 | -18 | 25.11 | 3.91 | 1.49 |
|   |            | 0.002 | 1.44E-05 | 0.02 | 0.0015 | 11 | 0.3 | 0.02 | 0.03 | 0.3 | 0.05 |   |   |
| 7 | 16482-3653 | -20.2 | 0.0038 | 4.0 | 0.0191 | 1390 | 168.8 | 20.6 | -25.4 | -22.5 | 19.4 | -20.99 | 106.08 |
|   |            | 0.3 | 0.0005 | 0.3 | 0.0005 | 90 | 0.8 | 0.3 | 0.9 | 0.9 | 0.4 |   |   |
| 8 | 16510-3731 | 6.2 | -3.4E-05 | -2 | -0.0001 | 17000 | 350.0 | 6.4 | 6.81 | 0.8 | 0.1 | 6.60 | -26.89 |
|   |            | 0.5 | 3.37E-05 | 4 | 0.0003 | 7000 | 40.0 | 1.1 | 0.07 | 0.6 | 0.3 |   |   |
| 9 | 17290-4358 | -6.9 | -0.00032 | -5 | 0.0004 | -17000 | 220.0 | 9 | -12.51 | 1.9 | 0.5 | -11.03 | -14.30 |
|   |            | 1.8 | 8.98E-05 | 6 | 0.000303 | 17000 | 30.0 | 4 | 0.18 | 0.6 | 0.5 |   |   |
| 10 | 17512-3033 | -9.0 | -0.00027 | 2.2 | -0.001 | -1500 | 170.0 | 9.3 | -9.43 | 0.5 | 1.1 | -9.55 | 39.47 |
|   |             | 0.3 | 8.3E-05 | 1.2 | 0.004 | 1200 | 7.0 | 0.4 | 0.17 | 0.7 | 0.3 |   |   |

Armed with the Rectilinear Elements, it is possible to give an ephemeris for the ρ and PA. This is gives in Table 7. Epochs are in the column headings.

*Table 7. Ephemeris for the 10 Sco Pairs based on the Rectilinear Motion.*

|   | WDS | 1991.25 PA° +/- | 1991.25 Sep" +/- | 2015.5 PA° +/- | 2015.5 Sep" +/- | 2020.0 PA° +/- | 2020.0 Sep" +/- | 2025.0 PA° +/- | 2025.0 Sep" +/- | 2030.0 PA° +/- | 2030.0 Sep" +/- | 2035.0 PA° +/- | 2035.0 Sep" +/- | 2040.0 PA° +/- | 2040.0 Sep" +/- |
|---|---|---|---|---|---|---|---|---|---|---|---|---|---|---|---|
| 1 | 16029-2501 | 345.59 | 4.53 | 343.85 | 4.48 | 343.52 | 4.47 | 343.15 | 4.46 | 342.79 | 4.46 | 342.42 | 4.45 | 342.05 | 4.44 |
|   |            | 0.09 | 0.00 | 0.00 | 0.00 | 0.02 | 0.00 | 0.04 | 0.00 | 0.05 | 0.00 | 0.07 | 0.00 | 0.09 | 0.00 |
| 2 | 16095-3239 | 83.99 | 7.68 | 83.77 | 7.65 | 83.73 | 7.65 | 83.69 | 7.64 | 83.64 | 7.64 | 83.60 | 7.63 | 83.55 | 7.62 |
|   |            | 0.00 | 0.00 | 0.00 | 0.00 | 0.00 | 0.00 | 0.00 | 0.00 | 0.00 | 0.00 | 0.00 | 0.00 | 0.01 | 0.00 |
| 3 | 16143-1025 | 152.81 | 22.29 | 153.18 | 22.35 | 153.24 | 22.36 | 153.32 | 22.37 | 153.39 | 22.38 | 153.47 | 22.39 | 153.54 | 22.40 |
|   |            | 0.12 | 0.05 | 0.00 | 0.00 | 0.01 | 0.00 | 0.02 | 0.00 | 0.08 | 0.03 | 0.10 | 0.04 | 0.13 | 0.05 |
| 4 | 16195-3054 | 318.52 | 23.32 | 318.35 | 23.46 | 318.32 | 23.49 | 318.28 | 23.52 | 318.25 | 23.55 | 318.21 | 23.58 | 318.18 | 23.61 |
|   |            | 0.07 | 0.03 | 0.00 | 0.00 | 0.02 | 0.01 | 0.03 | 0.01 | 0.04 | 0.02 | 0.06 | 0.02 | 0.07 | 0.03 |
| 5 | 16201-2003 | 332.83 | 46.66 | 332.81 | 46.64 | 332.81 | 46.63 | 332.80 | 46.63 | 332.80 | 46.62 | 332.79 | 46.62 | 332.79 | 46.62 |
|   |            | 0.00 | 0.00 | 0.00 | 0.00 | 0.00 | 0.00 | 0.00 | 0.00 | 0.00 | 0.00 | 0.00 | 0.00 | 0.00 | 0.00 |
| 6 | 16247-2942 | 356.55 | 4.65 | 359.13 | 4.07 | 359.69 | 3.97 | 0.35 | 3.85 | 1.05 | 3.73 | 1.79 | 3.62 | 2.58 | 3.50 |
|   |            | 0.04 | 0.00 | 0.00 | 0.00 | 0.01 | 0.00 | 0.02 | 0.00 | 0.03 | 0.00 | 0.05 | 0.00 | 0.06 | 0.00 |
| 7 | 16482-3653 | 139.32 | 23.65 | 138.33 | 23.89 | 138.15 | 23.93 | 137.95 | 23.98 | 137.75 | 24.03 | 137.56 | 24.08 | 137.36 | 24.13 |
|   |            | 0.03 | 0.01 | 0.00 | 0.00 | 0.01 | 0.00 | 0.01 | 0.00 | 0.02 | 0.01 | 0.02 | 0.01 | 0.03 | 0.01 |
| 8 | 16510-3731 | 4.66 | 6.76 | 4.63 | 6.76 | 4.63 | 6.76 | 4.62 | 6.76 | 4.62 | 6.76 | 4.61 | 6.76 | 4.60 | 6.76 |
|   |            | 0.06 | 0.00 | 0.00 | 0.00 | 0.01 | 0.00 | 0.02 | 0.00 | 0.04 | 0.00 | 0.05 | 0.00 | 0.06 | 0.00 |
| 9 | 17290-4358 | 168.22 | 13.43 | 168.18 | 13.44 | 168.18 | 13.44 | 168.17 | 13.44 | 168.16 | 13.45 | 168.16 | 13.45 | 168.15 | 13.45 |
|   |            | 0.03 | 0.00 | 0.00 | 0.00 | 0.01 | 0.00 | 0.01 | 0.00 | 0.02 | 0.00 | 0.03 | 0.00 | 0.03 | 0.00 |
| 10 | 17512-3033 | 189.58 | 10.10 | 189.72 | 10.11 | 189.75 | 10.12 | 189.78 | 10.12 | 189.81 | 10.12 | 189.84 | 10.12 | 189.87 | 10.12 |
|    |             | 0.05 | 0.00 | 0.00 | 0.00 | 0.01 | 0.00 | 0.02 | 0.00 | 0.03 | 0.00 | 0.04 | 0.00 | 0.05 | 0.00 |

## 8. Determination of the Orbit for Two Sco Pairs.

Following the technique presented in Letchford, White and Ernest, 2018b, it is possible to determine Grade 5 Orbital Elements for pairs that display very short arcs. For this analysis all historic data is considered as is the measure from Table 2.

The orbital elements for two Sco pairs are given in Table 8 and shown graphically in Appendix 3. Column headings in Table 8 are described in Letchford, White and Ernest, 2018b.

*Table 8. Orbital Elements for 2 Sco pairs.*

|   | WDS | P yrs | a " | I° | Ω° | T yr | e | ω° |
|---|---|---|---|---|---|---|---|---|
|   |   | +/- | +/- | +/- | +/- | +/- | +/- | +/- |
| 2 | 16095-3239 | 150000 | 40 | 100.0 | 120 | -11000 | 0.52 | 350 |
|   |   | 20000 | 4 | 2.3 | 7 | 15000 | 0.02 | 115 |
| 4 | 16195-3054 | 33000 | 34 | 118.0 | 110 | -1900 | 0.76 | 351 |
|   |   | 5000 | 4 | 1.2 | 10 | 100 | 0.09 | 20 |

Again, armed with these Orbital Elements, it is possible to give an ephemeris for the ρ and PA. These is gives in Table 9. Epochs are in the column headings. Units for ρ are arcseconds and units for PA are degrees. These predictions are in exact agreement with the rectilinear predictions of Table 7.

*Table 9. Ephemeris for the 2 Sco Pairs based on the Orbital Motion.*

|   | WDS | 1991.25 PA° | 1991.25 Sep" | 2015.5 PA° | 2015.5 Sep" | 2020.0 PA° | 2020.0 Sep" | 2025.0 PA° | 2025.0 Sep" | 2030.0 PA° | 2030.0 Sep" | 2035.0 PA° | 2035.0 Sep" | 2040.0 PA° | 2040.0 Sep" |
|---|---|---|---|---|---|---|---|---|---|---|---|---|---|---|---|
| 2 | 16095-3239 | 84.00 | 7.68 | 83.78 | 7.65 | 83.74 | 7.65 | 83.69 | 7.64 | 83.65 | 7.64 | 83.60 | 7.63 | 83.55 | 7.62 |
| 4 | 16195-3054 | 318.53 | 23.31 | 318.36 | 23.46 | 318.32 | 23.49 | 318.29 | 23.52 | 318.25 | 23.55 | 318.22 | 23.58 | 318.18 | 23.61 |

## 9.    Conclusion.

We presented measures for 10 Pairs in the constellation of Scorpius using a C14 telescope, lucky imaging, drift scans and the *Reduc* software. The separations of α Cen AB, as determined from the orbital elements of Pourbaix and Boffin (2016) were used as the image scale and position angle calibrator, where PA calibration was undertaken using drift scans.

Our internal uncertainties are ~0.06 arcseconds in ρ and ~0.06 degree in PA. There is excellent agreement with historic data extrapolated to epoch of observation (~2018.53), and micro-arcsecond positions from the GAIA database where the differences were ~0.05 arcsecond in ρ and ~0.15 degrees in PA.

There is excellent agreement between the extrapolated historic observations and these from GAIA.

In addition, we presented rectilinear elements for 10 Sco pairs and Orbital Elements for two of them. Ephemera are given for these pairs based on both the rectilinear elements are the orbital elements.

**Acknowledgements.**

We acknowledge the use of the following resources:

- SIMBAD Astronomical Database, operated at CDS, Strasbourg, France, https://simbad.u-strasbg.fr/simbad
- The *Aladin sky atlas* developed at CDS, Strasbourg Observatory, France, https://aladin.u-strasbg.fr
- The *Washington Double Star Catalog* maintained by the USNO. (WDS), https://ad.usno.navy.mil/wds

- All-sky Compiled Catalogue of 2.5 million stars, 3rd version (ASCC), http://vizier.u-strasbg.fr/viz-bin/VizieR-3?-source=I/280B/ascc
- The Gaia Catalogue (Gaia DR2, Gaia Collaboration, 2018), from VizieR (GAIA DR2), http://vizier.u-strasbg.fr/viz-bin/VizieR-3?-source=I/345/gaia2
- *SharpCap* astrophotography software developed by Robin Glover, https://www.sharpcap.co.uk

In addition, we thank Jenny Stevens for her support for author Meg Emery, and Florent Losse for the Reduc software.

# Appendices.

## Appendix 1 – Trends Shown by the Incorporation of Historic Data.

Historic positional measures have been obtained from supplementary catalogues of the WDS. For the 10 pairs a total of 420 observations are available dating from 1783.23.

This Appendix presents the $\rho$ and PA for the 10 Sco pairs. All plots of PA (and $\rho$) are at the epoch and equinox of observation. Data points in orange have been rejected from the trend. The green points are the measures from this work (from Table 2).

A trend towards better quality data is visible as the spread of data points around the trend line is converging with epoch.

Each plot has been fitted by an unweighted linear trend line and the fitted parameters are given in Table 3 along with the derived correlation coefficient, $R^2$.

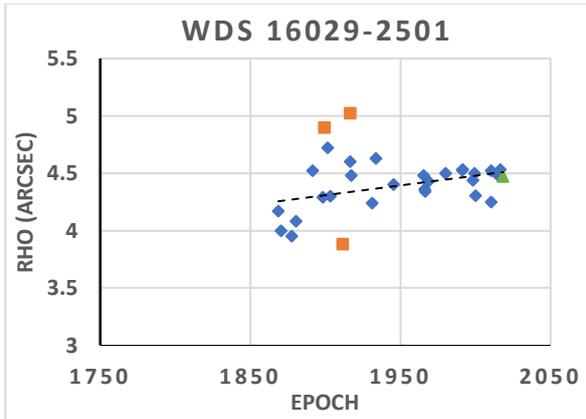
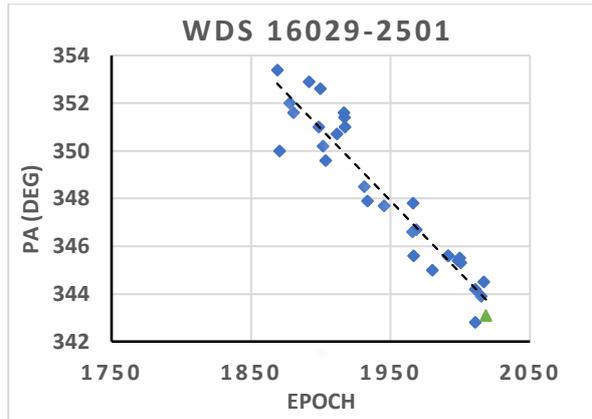
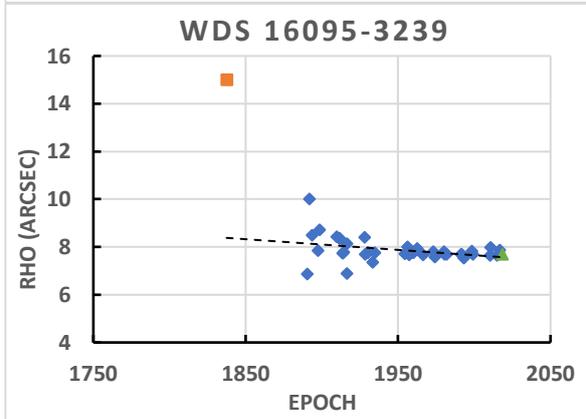
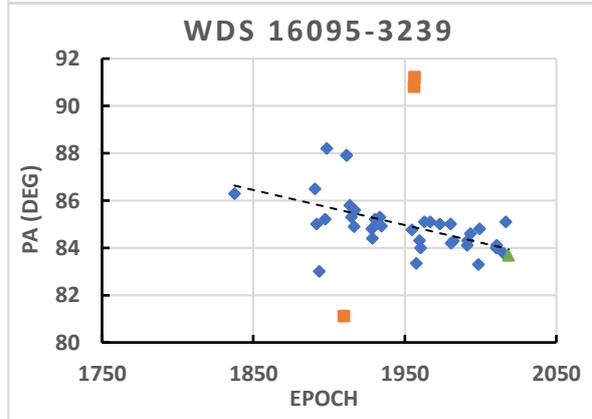
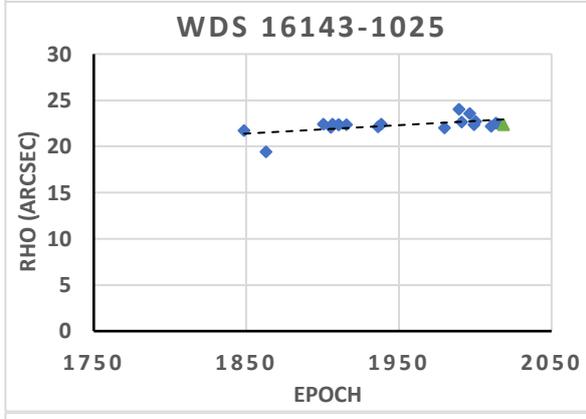
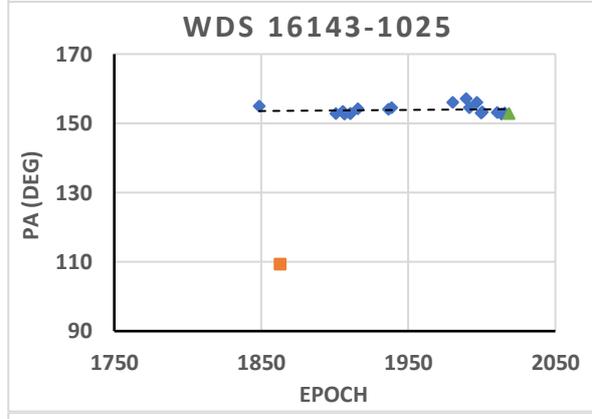
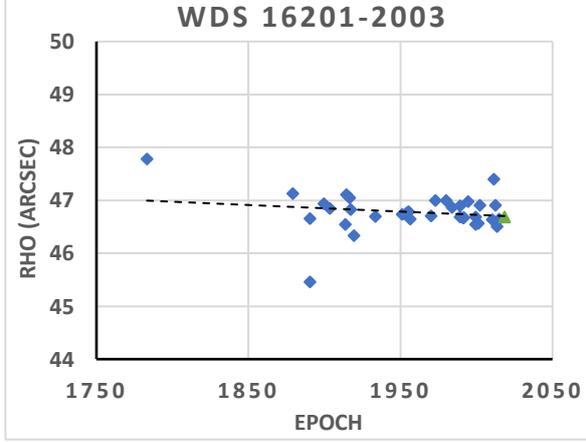
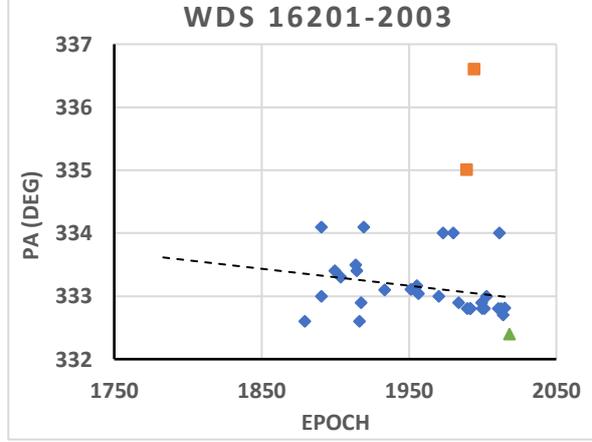

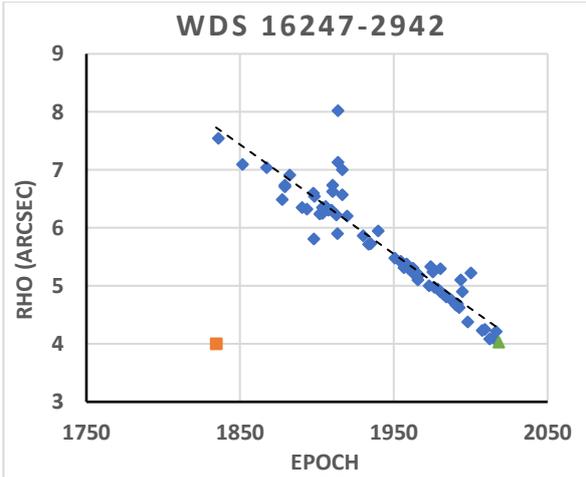
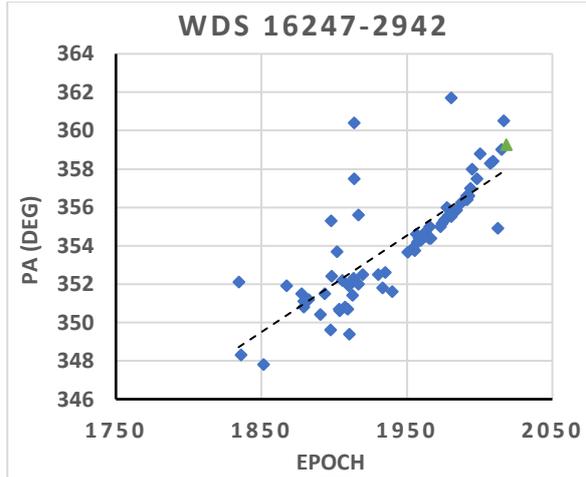
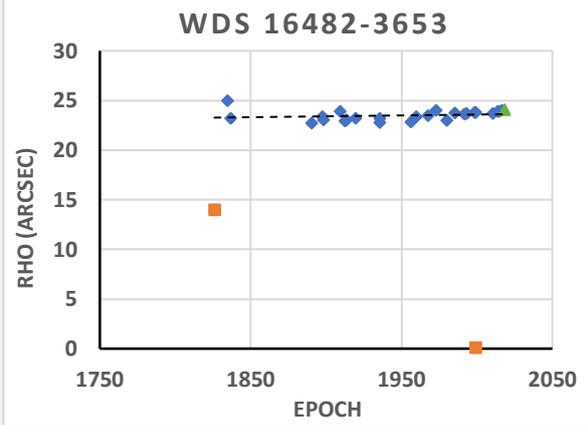
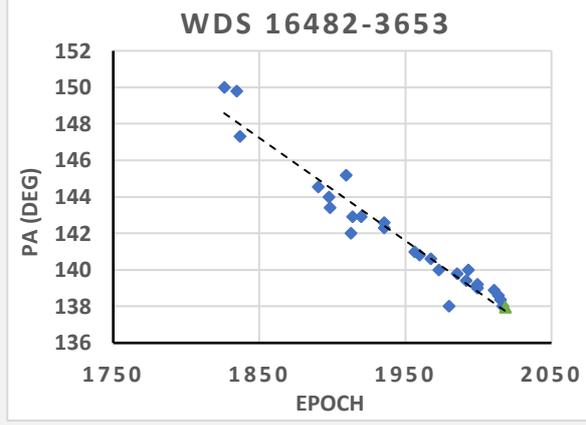
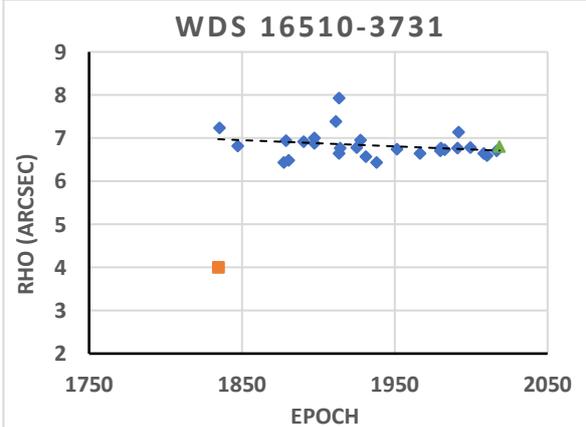
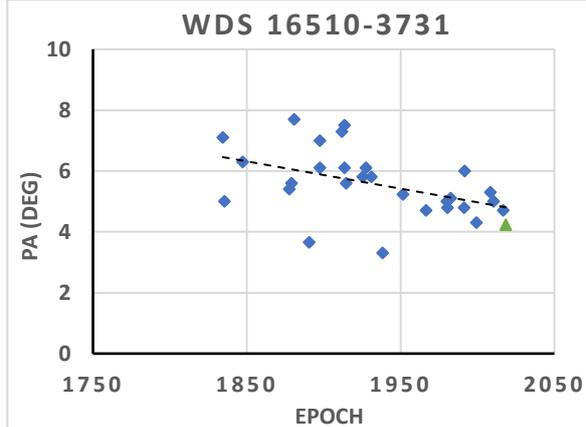
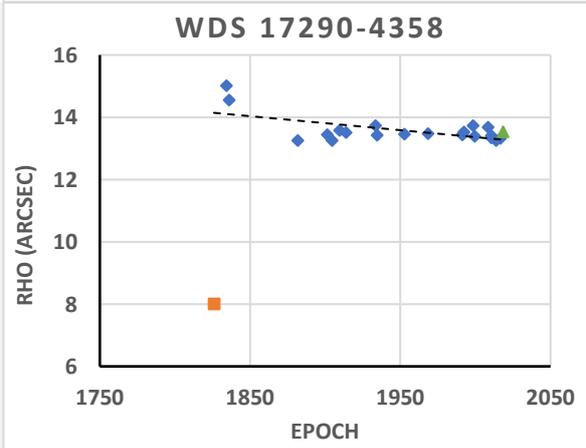
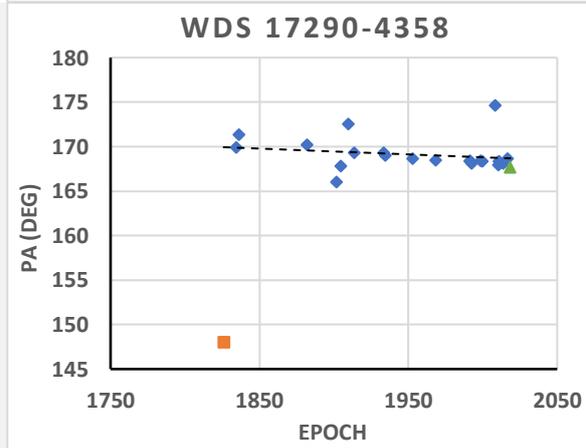

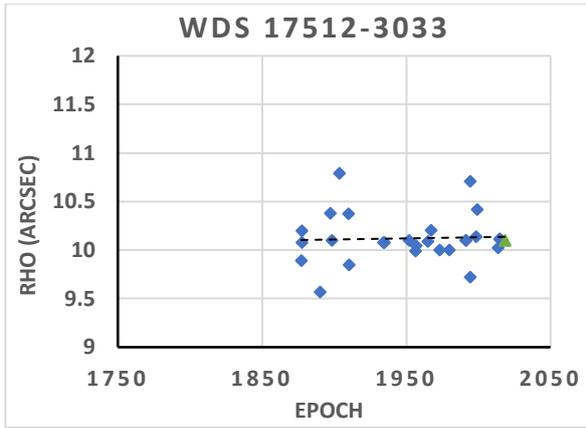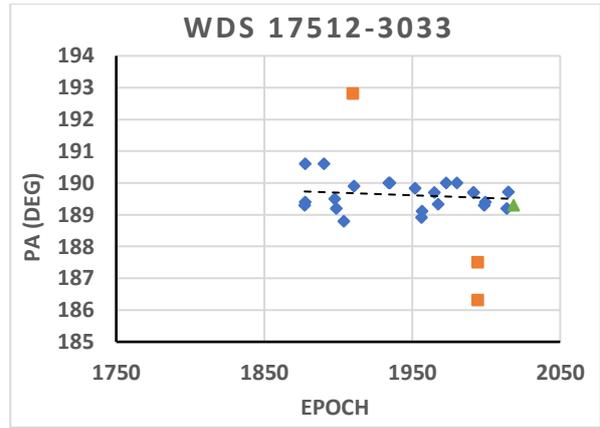

# Appendix 2 – Rectilinear Motion of the Ten Sco Pairs.

Plots of the rectilinear motion of the 10 Sco doubles are given here. Left hand figures are unzoomed – right hand are zoomed. Historical data from the WDS have been incorporated and their position angles have been precessed from Equinox of date to Equinox J2000.0 using proper motions. The WDS data for 1991.25 (HIP – from the HIPPARCOS mission) are already at Equinox J2000.0. Precessed WDS observations are represented in the plots by a '+'.

The HIP and GAIA positions are represented by a red circle and green square respectively. The dotted ellipses are the uncertainty ellipses for the t0 (un-zoomed figure for each pair). If they cannot be seen in the plots, it is because of the plot scale. Uncertainty ellipses for the HIP and GAIA were also plotted but in each case they too may be too small to see at the scales that are needed to represent all relevant data.

Red line is HIP proper motion and Green is GAIA. The black line is the rectilinear motion based on the HIP and GAIA position. Rectilinear Elements are given in Table 6 and projected ρ and PA in Table 7.

For additional understanding into the this process, read in Letchford, White and Ernest, 2018b.

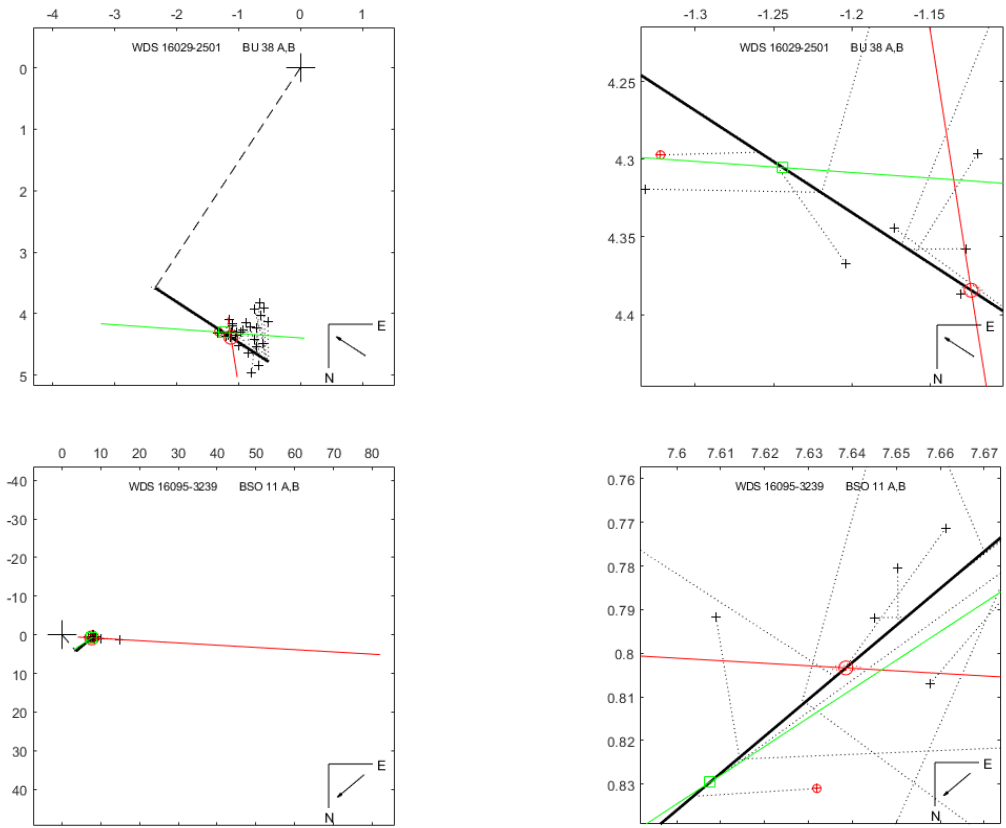

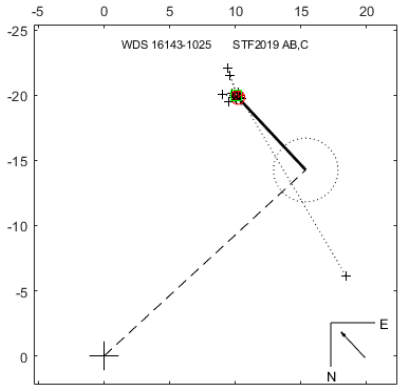
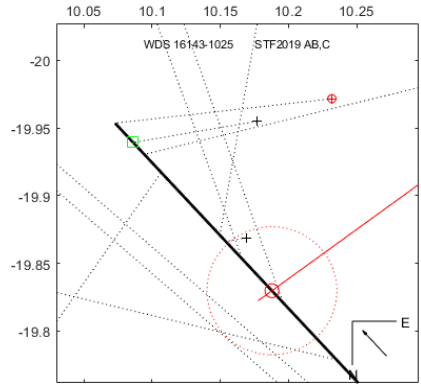
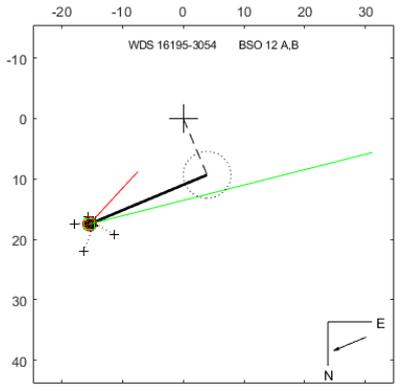
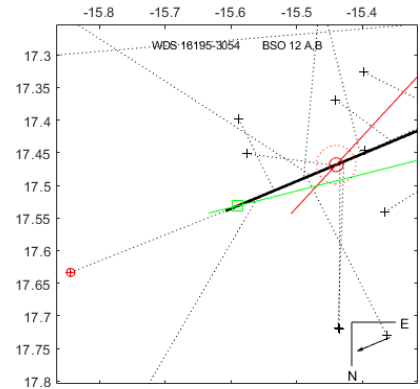
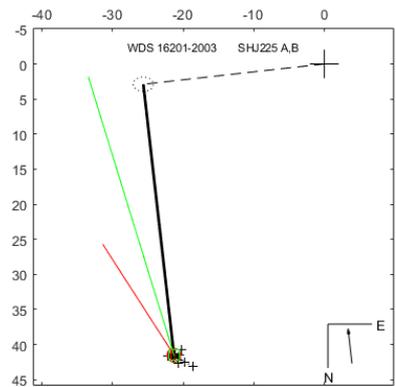
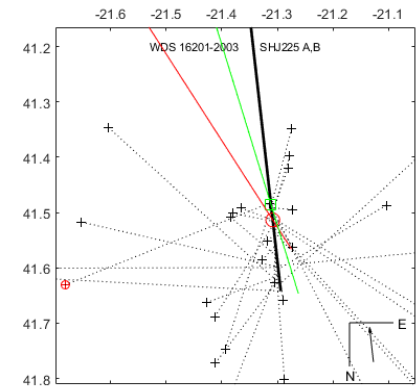
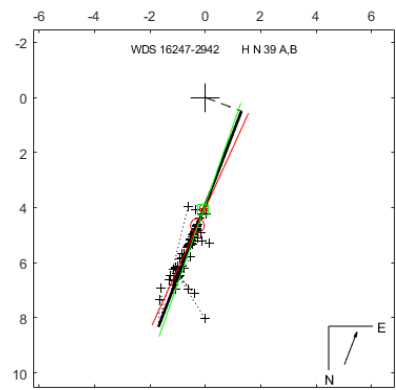
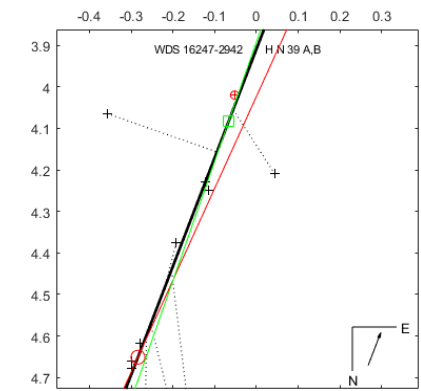

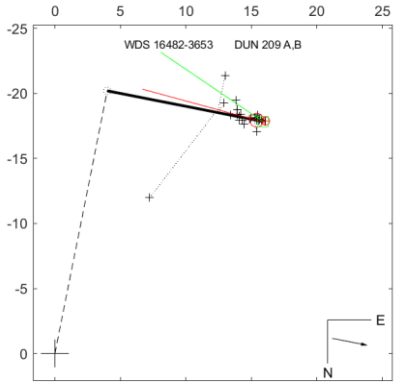
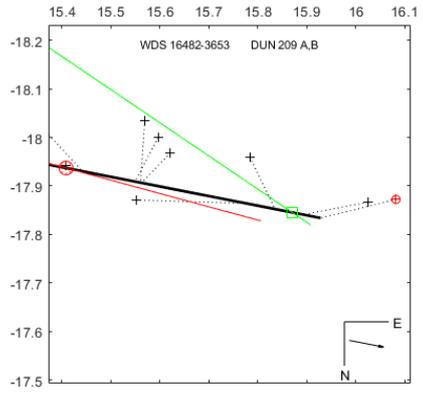
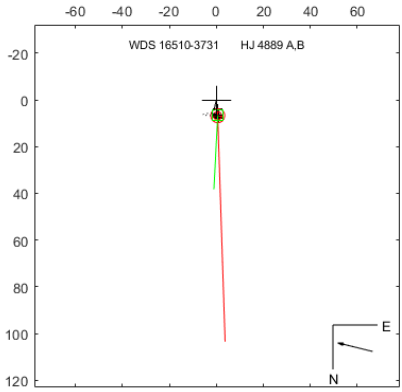
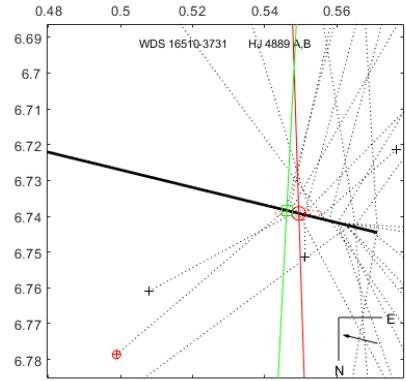
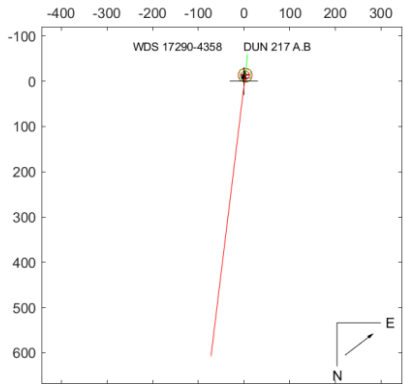
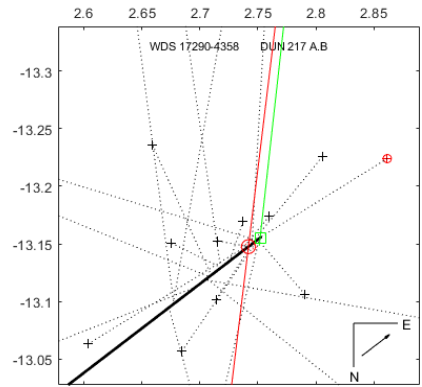
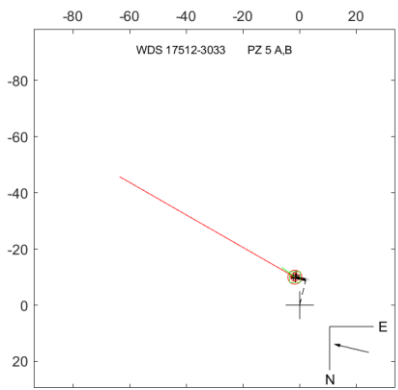
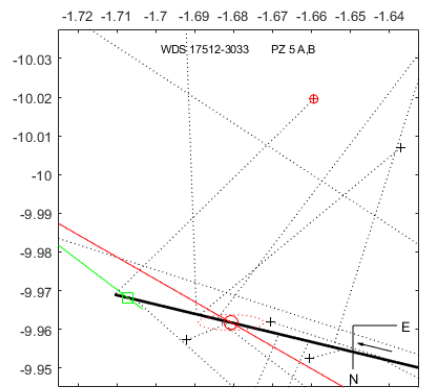

# Appendix 3 – Orbits Found for the Two Sco Pairs.

The family of orbits for 2 Sco pairs. The best orbit (smallest residuals from historic data) is bolded and the elements are in Table 8. Predicted ρ and PA based on the Orbital Elements are given in Table 9.

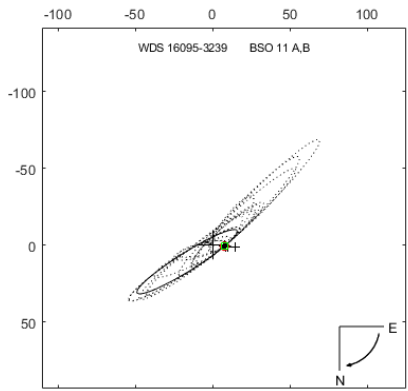 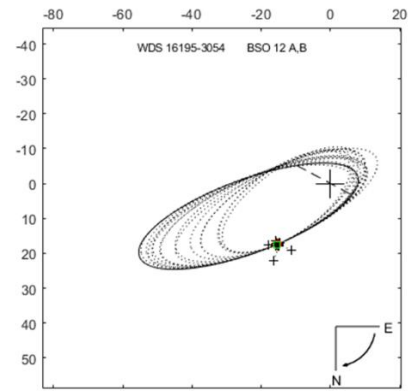